\documentclass{article}

    \PassOptionsToPackage{numbers, compress}{natbib}
    \usepackage[preprint]{neurips_2024}

\usepackage[utf8]{inputenc} % allow utf-8 input
\usepackage[T1]{fontenc}    % use 8-bit T1 fonts
\usepackage{hyperref}       % hyperlinks
\usepackage{url}            % simple URL typesetting
\usepackage{booktabs}       % professional-quality tables
\usepackage{amsfonts}       % blackboard math symbols
\usepackage{nicefrac}       % compact symbols for 1/2, etc.
\usepackage{microtype}      % microtypography
\usepackage{xcolor}         % colors
\usepackage{xspace}
\usepackage[pdftex]{graphicx}
\usepackage{subcaption}
\usepackage{makecell}
\usepackage[acronym]{glossaries} 
\robustify{\gls}% Make \gls not fragile
\newacronym{DNN}{DNN}{Deep Neural Networks}
\newacronym{DL}{DL}{Deep Learning}
\newacronym{CNN}{CNN}{Convolutional Neural Network}
\newacronym{ML}{ML}{Machine Learning}
\newacronym{NN}{NN}{Neural Network}
\newacronym{ViT}{ViT}{Vision Transformer}
\newacronym{MHA}{MHA}{Multi-head Attention}
\newacronym{MLP}{MLP}{Multilayer Perceptron}
\newacronym{NLP}{NLP}{Natural Language Processing}
\newacronym{IoT}{IoT}{Internet of Things}
\newacronym{EMA}{EMA}{Exponential Moving Average}
\newacronym{MCU}{MCU}{Microcontroller}
\newacronym{TFLite}{TFLite}{TensorFlow Lite}
\newacronym{RDO}{RDO}{Rate Distortion Optimization}
\newacronym{CABAC}{CABAC}{Context Adaptive Binary Arithmetic Coding}
\newacronym{DSP}{DSP}{Digital Signal Processor}
\newacronym{DCT}{DCT}{Discrete Cosine Transform}
\newacronym{GDN}{GDN}{Generalized Divisive Normalization}
\newacronym{M-JPEG}{M-JPEG}{Motion-JPEG}

\usepackage{booktabs}
\usepackage{multirow}
\usepackage{graphicx,wrapfig,lipsum}
\usepackage{algorithm2e}
\usepackage{algpseudocode}
\usepackage{mathtools}
\usepackage{nicematrix}
\usepackage{booktabs}
\usepackage{pifont}% http://ctan.org/pkg/pifont
\newcommand{\name}{\texttt{MCUCoder}\xspace}

\title{\name: Adaptive Bitrate Learned Video Compression for IoT Devices}
\author{%
  Ali Hojjat, Janek Haberer, Olaf Landsiedel \\
  Kiel University, Germany\\
  \texttt{\{ali.hojjat, janek.haberer, olaf.landsiedel\}@cs.uni-kiel.de} \\
}

\begin{document}

\maketitle

\begin{abstract}
The rapid growth of camera-based \gls{IoT} devices demands the need for efficient video compression, particularly for edge applications where devices face hardware constraints, often with only 1 or 2 MB of RAM and unstable internet connections. Traditional and deep video compression methods are designed for high-end hardware, exceeding the capabilities of these constrained devices. Consequently, video compression in these scenarios is often limited to \gls{M-JPEG} due to its high hardware efficiency and low complexity. This paper introduces \name, an open-source adaptive bitrate video compression model tailored for resource-limited IoT settings. \name features an ultra-lightweight encoder with only 10.5K parameters and a minimal 350KB memory footprint, making it well-suited for edge devices and \glspl{MCU}. While \name uses a similar amount of energy as  \gls{M-JPEG}, it reduces bitrate by 55.65\% on the MCL-JCV dataset and 55.59\% on the UVG dataset, measured in MS-SSIM. Moreover, \name supports adaptive bitrate streaming by generating a latent representation that is sorted by importance, allowing transmission based on available bandwidth. This ensures smooth real-time video transmission even under fluctuating network conditions on low-resource devices.
% Source code available at \url{https://github.com/ds-kiel/mcucoder}
Source code available at \url{https://github.com/ds-kiel/MCUCoder}.
\end{abstract}

\begin{figure}[h]
    \centering
    \begin{subfigure}[b]{1.\textwidth}
        \centering
        \includegraphics[trim=15 7 0 5, clip, width=1.0\textwidth]{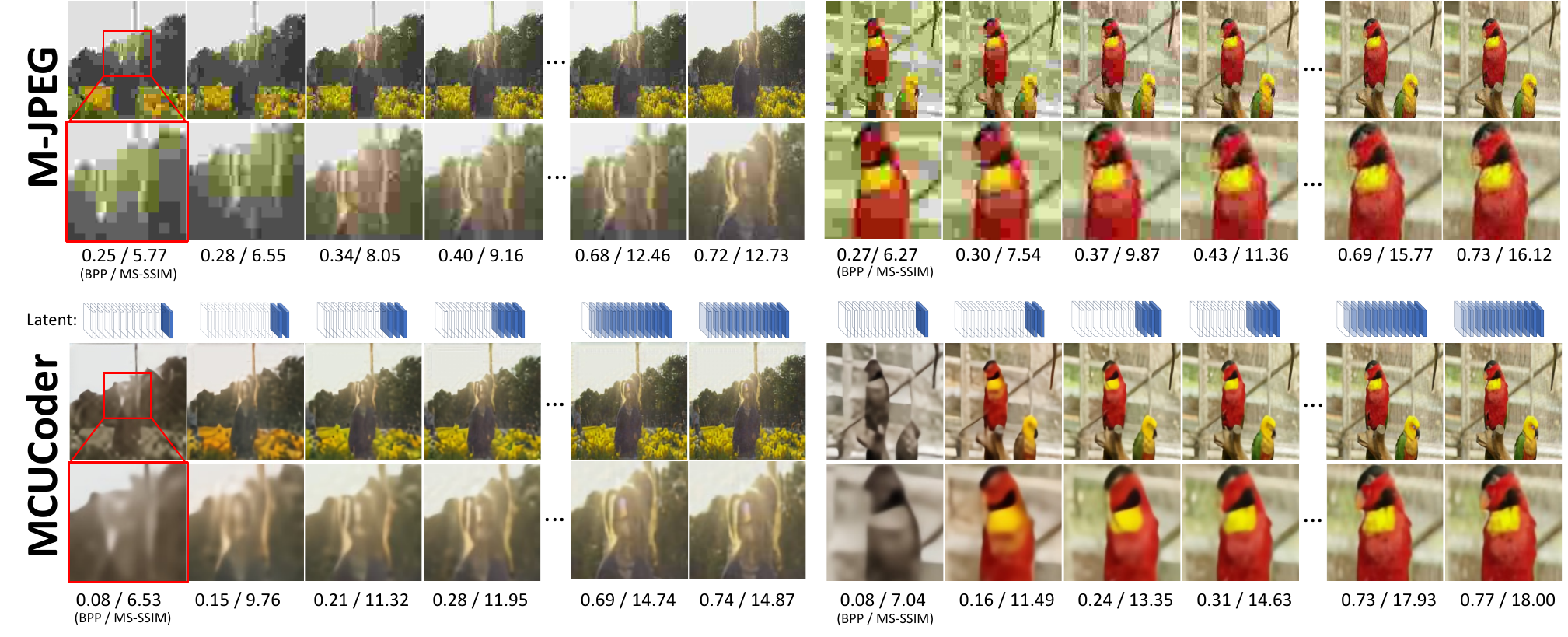}
    \end{subfigure}
    \caption{Qualitative comparison of \name and \gls{M-JPEG}  across various compression rates on two videos from the MCL-JCV \cite{mcl_jcv} and UVG \cite{mercat2020uvg} datasets. As we can see, \name offers a significantly better MS-SSIM/bpp trade-off. For instance, at 0.15 bpp in the left example, with \name we can see the person's face whereas with \gls{M-JPEG} we need at least 0.34 bpp to make out the face.  Note that the images in each column do not necessarily have the same bitrate. More examples are reported in Appendix \ref{appenix_D}.}  
\label{fig:demo}
\vspace{-2em}

\end{figure}

\section{Introduction}
% \vspace{-1.3em}

\textbf{Motivation:}
\begin{wrapfigure}{L}{5cm}
\includegraphics[trim=20 10 20 20, width=4.8cm]{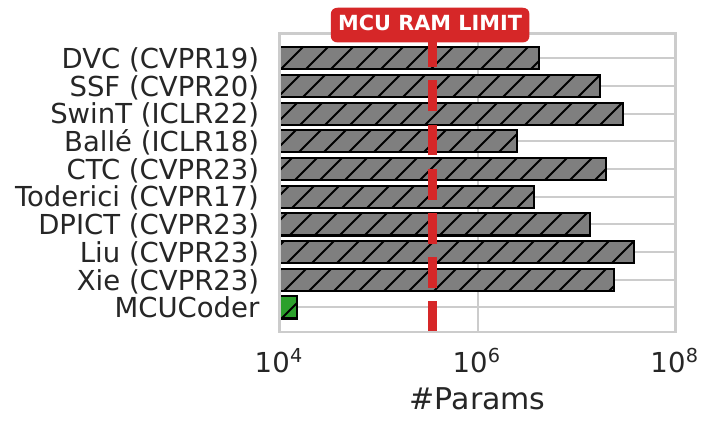}
% \captionsetup{font=tiny}
\caption{Number of parameters of \name and other learned image compression \cite{balle2018variational, toderici2017full, lee2022dpict, jeon2023context, Liu_2023_CVPR, xie2021enhanced, zhu2022transformer} and video compression models \cite{agustsson2020scale, lu2019dvc}.}
\vspace{-1em}
\label{fig:params}
\end{wrapfigure}
The number of camera-based \glspl{IoT} devices using always-on \gls{MCU} is growing rapidly, reaching tens of billions \cite{lin2020mcunet}. These devices are widely used in applications such as surveillance cameras \cite{hu2020starfish, josephson2019wireless, naderiparizi2018towards}, wearable cameras \cite{veluri2023neuricam}, robotics \cite{nakanoya2023co}, wildlife monitoring \cite{iyer2020wireless}, road monitoring \cite{limitnet}, and smart farming \cite{koh2021wilds}. Typically, they capture raw frames through a camera sensor, encode them, and transmit the compressed version to a server via the Internet for further processing, including human observation or AI tasks such as object detection and classification \cite{limitnet}. Therefore, a video encoder is necessary to efficiently compress the captured frames before transmission. However, in \gls{IoT} environments, there are two primary limitations: constrained hardware resources and limited communication bandwidth.

\textbf{1 - Limited Hardware:} Although traditional video codecs like H.264 \cite{h264}, H.265 \cite{h265}, and the newer H.266 \cite{h266} provide excellent performance, they demand significant hardware for extracting the intra and inter-frame correlations. For example, H.265 encoding involves highly computationally intensive tasks such as motion estimation with sub-pixel accuracy, \gls{RDO} for choosing optimal intra-prediction modes, and \gls{CABAC} for entropy coding.
Additionally, a single video frame at $224 \times 224$ resolution requires about 150 KB of RAM, which is a lot for the low-cost, low-energy \glspl{MCU} used in \gls{IoT} devices that typically have only 1–2 MB of RAM.
Consequently, inter-frame compression or any other kind of multi-frame analysis is not practically feasible on such constrained devices. Similarly, while \glspl{NN} and AI-based compression methods outperform traditional models \cite{agustsson2020scale, lu2019dvc}, they also often require considerable RAM and GPU resources.
For instance, just storing a model with 1M parameters requires around 4 MB of RAM; see Fig.~\ref{fig:params}. 
% Furthermore, adapting these models for deployment on \glspl{MCU} with frameworks like TFLite Micro imposes limitations, as they mainly support basic operations such as \glspl{MLP} and Convolutions with specific kernel size and strides \cite{hu2020starfish}. In contrast, modern \gls{NN} compression models often use more complex components like attention layers, \gls{GDN} \cite{balle2015density}, and non-local blocks \cite{Wang_2018_CVPR}.
% These requirements are often too demanding for such a constrained \glspl{MCU}; see Fig.\ref{fig:params}.
As a result, in such settings, devices are typically limited to using \gls{M-JPEG} \cite{jpeg}, a video compression format where each frame is compressed individually as a JPEG image, which is efficient and hardware-friendly.

\textbf{2 - Limited Internet:} Many IoT devices are located in remote areas where Internet connection is weak and unstable, making it necessary for the encoder to have an \textbf{Adaptive Bitrate Encoding} that can generate video streams with varying bitrate. This feature allows the encoder to dynamically adjust its quality according to the available bandwidth, ensuring continuous and smooth playback. This is especially important for real-time applications like live monitoring, where it is crucial to avoid interruptions and maintain a consistent user experience despite fluctuating network conditions. However, implementing an adaptive bitrate encoder adds complexity, as it requires mechanisms to prioritize bit stream information based on its impact on frame quality (e.g., PSNR or MS-SSIM), which is challenging for constrained devices.

\textbf{Approach:} To address these challenges, we introduce \name, \textit{an adaptive bitrate deep video compression model tailored for resource-limited IoT devices}. Our approach focuses on creating an "asymmetric" compression model that features an ultra-lightweight encoder designed to be both computationally efficient and memory-friendly.
Also, \name produces an "adaptive bitrate" bitstream. Specifically, in \name, we train the encoder using stochastic dropout such that, instead of explicitly detecting the important parts, it produces latent channels that are sorted based on importance.
%For example, if the encoder produces \( N \) feature maps, the initial feature maps are more important than the subsequent ones.
Afterward, based on the available internet bandwidth, the encoder transmits the first $k$ channels to the decoder; see Fig.~\ref{fig:demo}. 
This approach is beneficial for low-power MCUs since it shifts the complexity of identifying important data to the training phase rather than the inference phase. Also, by employing stochastic dropout training, the decoder can reconstruct the frame even with partial data availability, which is essential for maintaining smooth and uninterrupted video transmission in real-time applications, where network conditions can vary.
Additionally, \name's encoder is INT8 quantized, allowing it to utilize \gls{DSP} and CMSIS-NN \cite{CMSIS-NN} accelerators for faster processing and reduced power consumption.

\newpage
\textbf{Contributions:}
\vspace{-0.75em}
\itemsep0em
\begin{enumerate}
\itemsep0em

    \item \name has an ultra-lightweight encoder with only 10.5K parameters and a minimal memory footprint of roughly 350KB RAM on nRF5340 and STM32F7 \glspl{MCU}, making it suitable for such low-resource IoT devices.
    
    \item \name has an energy-efficient INT8 quantized encoder, which leverages the \gls{MCU}'s \gls{DSP} and CMSIS-NN accelerators to achieve JPEG-level energy efficiency. Compared to its main baseline, \gls{M-JPEG}, it saves 55.65\% overall bit rate on the MCL-JCV dataset and 55.59\% on the UVG dataset, measured in MS-SSIM.
    
    \item \name produces a progressive bitstream that enables adaptive bitrate streaming, allowing robust video transmission under varying network conditions.
\end{enumerate}
\vspace{-1em}

\section{Related Work}
\vspace{-.5em}

\begin{figure*}
    \centering
    \begin{subfigure}[b]{1.\textwidth}
        \centering
        \includegraphics[trim=0 0 0 0, clip, width=1\textwidth]{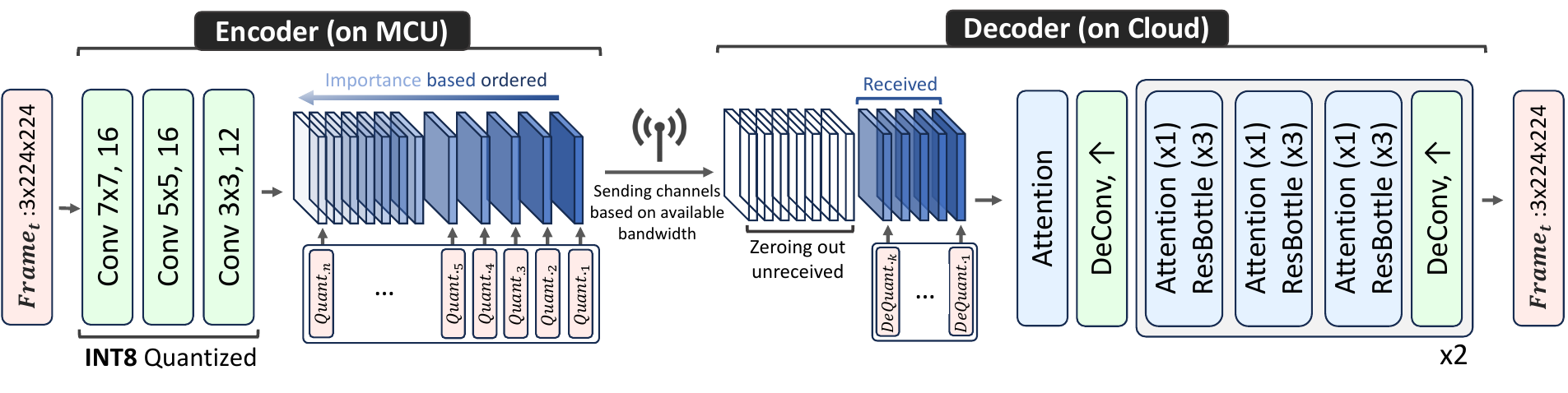}
    \end{subfigure}
    \caption{Overview of \name architecture. The encoder compresses the input frame into a sorted latent space. Afterward, channels are independently quantized and transmitted based on available bandwidth. The decoder reconstructs the frame by zeroing out missing channels.}   
\label{fig:ARCH}
\vspace{-2em}
\end{figure*}

\label{sec:rw}
\textbf{Traditional and \gls{NN} based video compression}: Video compression is a field that has been evolving for decades. Beyond traditional codecs like H.264 \cite{h264}, H.265 \cite{h265}, and H.266 \cite{h266}, deep learning-based approaches often replace conventional modules such as motion compensation \cite{agustsson2020scale, yang2020learning}, transform coding \cite{zhu2022transformer, gao2021neural}, and entropy coding \cite{xiang2023mimt, mentzer2022vct}.
Also, some work has been done regarding the end-to-end optimization of video compression models \cite{he2020video, van2021instance, khani2021efficient}. Lu et al. \cite{lu2019dvc} introduce DVC, the first end-to-end deep video compression model. Hu et al. \cite{hu2022coarse, hu2021fvc} extend DVC to operate in both pixel and feature domains.
Li et al.\cite{li2021deep} and Lie et al. \cite{liu2020conditional} reduce bitrates by modeling probabilities over video frames using conditional coding. Also, in recent years, there has been growing interested in using implicit neural representations for video compression \cite{kwan2024hinerv, chen2021nerv}.
However, due to their substantial hardware requirements, these models are unsuitable for deployment on low-resource \gls{IoT} devices.

\textbf{Video compression for \gls{IoT}:} 
We can categorize \gls{IoT}-based video encoders into two parts: hardware-based and software-based.
Hardware approaches primarily focus on designing more power-efficient camera sensors \cite{morishita2021cmos, ji2016220pj, bejarano2022millimeter} and more efficient \gls{MCU} circuits and processors \cite{lefebvre20217, rossi20214, xu2020macsen}.
Due to its simplicity, scalability, low latency, and very low energy consumption, the most common software-based video encoder on \gls{IoT} devices is \gls{M-JPEG} \cite{jpeg}.
Nevertheless, there have been few works exploring alternative software-based models: Veluri et al. \cite{veluri2023neuricam} employ \gls{M-JPEG} on the encoder to capture black-and-white and colorized frames at two different resolutions and uses super-resolution methods to interpolate and colorize frames on the decoder. However, unlike \name, it is not adaptive and relies on a JPEG encoder on \glspl{MCU}.
Hu et al.\cite{hu2020starfish} propose a deep image encoder model for \glspl{MCU}, but it is also non-adaptive. Additionally, they patchify the input, which significantly increases encoding time, making it impractical for real-time video compression.
\name combines the advantages of both worlds: it offers the adaptive bitrate feature of more complex encoders, while maintaining the efficiency necessary for low-resource devices, making it an ideal solution for \gls{IoT} video compression.

\section{\name}\label{sec:design}
\vspace{-.5em}
In this section, we introduce \name, an adaptive bitrate asymmetric video compression model, specifically designed for IoT settings. We begin by detailing the asymmetric encoder-decoder architecture of \name, including the customized quantization processes. Then, we present the stochastic dropout training method, which trains the encoder of \name to store information in its channels based on importance.

\begin{figure*}
    \centering
    % First minipage
    \begin{minipage}[t]{0.45\textwidth} % Use [t] for top alignment
        \centering
        \includegraphics[trim=0 0 0 0, clip, width=.7\textwidth]{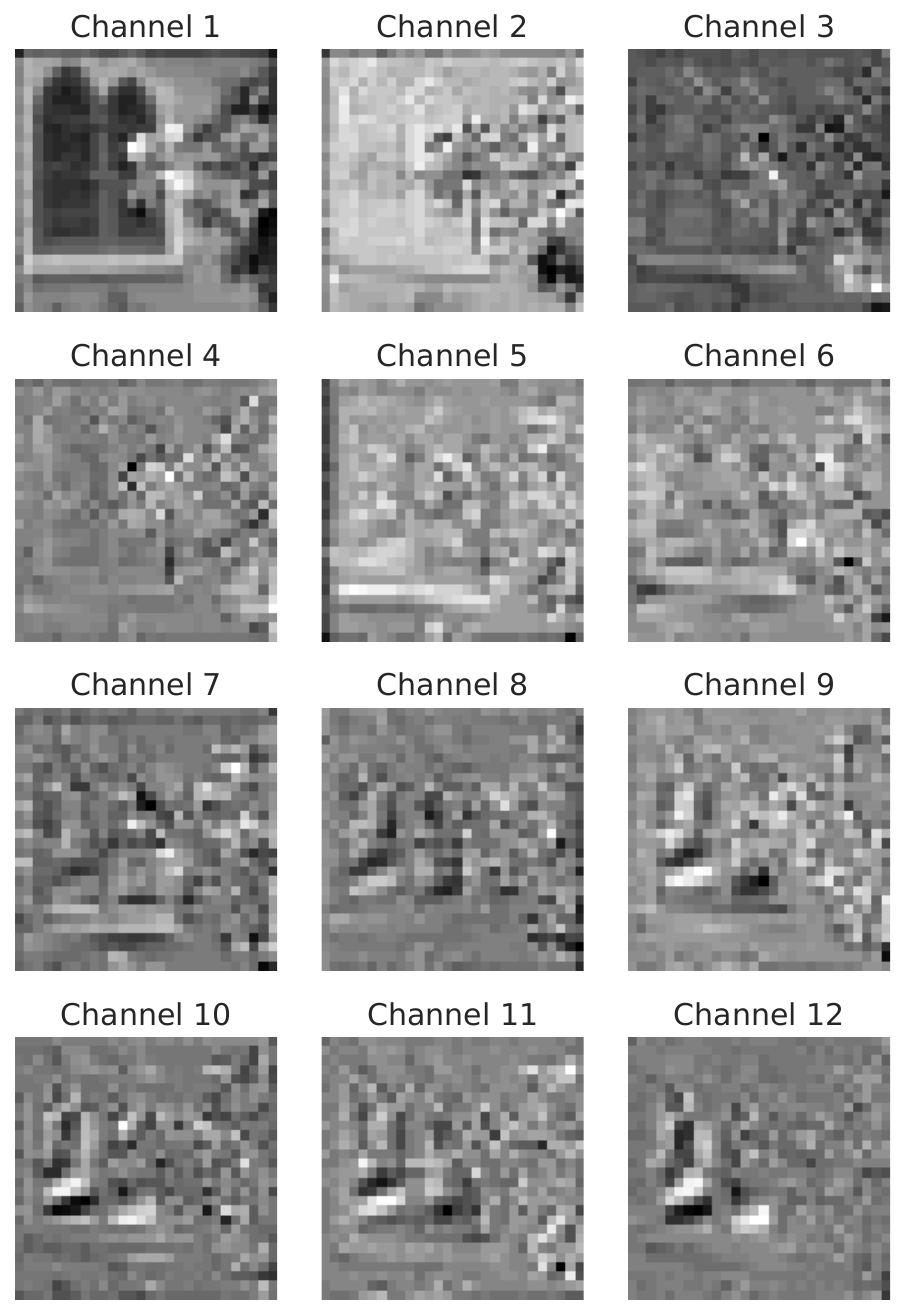}
        \caption{\name latent channels: Early channels (important ones) capture low-frequency features, while later channels capture high-frequency features, similar to the DCT in JPEG.}  % Add a caption if needed
        \label{fig:latent}
    \end{minipage}
    \hfill
    % Second minipage
    \begin{minipage}[t]{0.5\textwidth} % Use [t] for top alignment
        \centering
        \includegraphics[trim=0 0 0 0, clip, width=.7\textwidth]{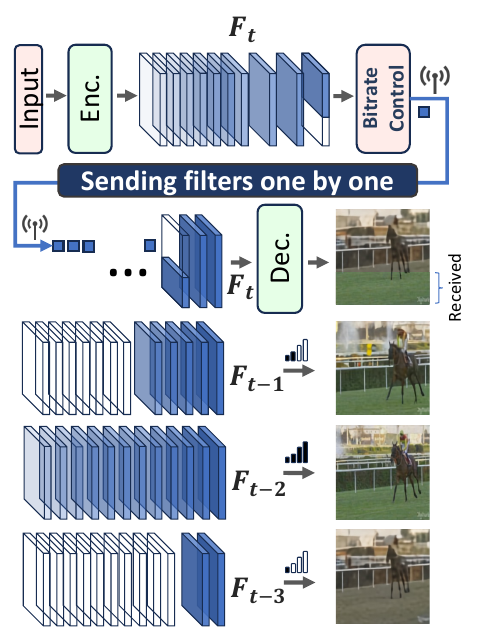}
        \caption{An example of \name bitrate adaptation under dynamic network bandwidth, where the bitrate control module acts as a gate to determine the number of channels to send.} 
        \label{fig:example}
    \end{minipage}
  
    \label{fig:combined_figure}
    \vspace{-1em}
\vspace{-0.5em}
\end{figure*}

\begin{wrapfigure}{R}{2.7cm}
\includegraphics[trim=10 10 20 20, width=2.7cm]{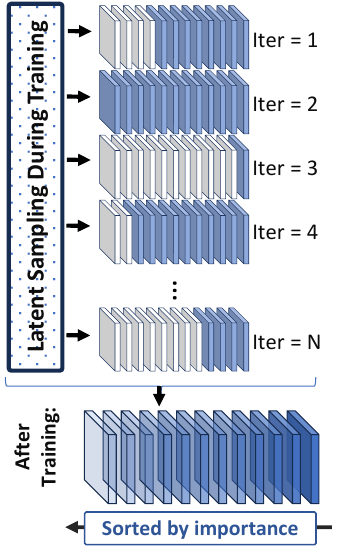}
\caption{Stochastic dropout training}
\vspace{-1em}
\label{fig:training}
\end{wrapfigure}

\textbf{Asymmetric Compression:}
\glspl{MCU} are characterized by highly constrained hardware resources, such as limited RAM, CPU, FLASH, and power availability. Additionally, existing \gls{MCU}-specific \gls{NN} frameworks like TFLite Micro support only a limited set of \gls{NN} layers \cite{hu2020starfish}. To address these constraints, we propose an asymmetric \cite{yao2020deep} encoder-decoder architecture optimized for constrained devices.
Due to hardware constraints, \name encodes each frame independently, as inter-frame compression is not feasible.
The encoder contains only 10.5K parameters, while the decoder utilizes approximately 3M parameters and leverages SOTA image decompression blocks; see Fig.~\ref{fig:ARCH}.
The encoding process begins by passing input frame $f_{t}$ through three convolutional layers.
To maximize the data range for subsequent quantization, no activation function is applied in the final encoder layer, avoiding the negative truncation caused by ReLU. Afterward, each channel of the latent is quantized into INT8 individually, followed by a further reduction to 5-bit precision to enhance compression efficiency. For the decoder, inspired by \cite{he2022elic}, we integrate a combination of attention blocks \cite{cheng2020learned} and residual bottleneck blocks \cite{he2016deep} to reconstruct the frame; see Fig. \ref{fig:ARCH}.

\textbf{Stochastic dropout training:}
Bitrate adaptation is a feature that typically introduces additional complexity to the encoding process, which can be challenging to implement on \glspl{MCU} due to resource constraints.
In the literature, dropout \cite{srivastava2014dropout} serves as a powerful tool for enhancing generalization in \glspl{NN}. Building on this insight, we employ a "biased" version of dropout to train \name in a way that instead of random dropping, it drops from the tail of the latent \cite{Hojjat_2023_CVPR, haberer2024hydravit}. 
Specifically, on each iteration, after the encoder $E$ gets the input frame $f_t$, it generates the latent representation $z_{N}$, where $N$ is the number of the channels of the latent. Afterward, from a uniform distribution, denoted as $\mathcal{U}_{(0, 1)}$, it generates a number, denoted as $k$, and drops (zero out) the last $\lfloor k \times N \rfloor $ channels from  $z_{N}$. As a result, instead of $z_{N}$, the decoder $D$ gets $z_{[0:\lfloor k \times N \rfloor]}$, fills the missing channels with zero, and then reconstructs the output.
\begin{equation}
  f_{t}\xrightarrow{} E(f_{t}) \xrightarrow{} z_{N} \xrightarrow[]{k \sim  \mathcal{U}_{(0, 1)}}z_{[0:\lfloor k \times N \rfloor]} \xrightarrow{} D(z_{[0:\lfloor k \times N \rfloor]}) \xrightarrow{} \hat{f}_{t}
\label{eq:1}
\end{equation}
This tailored version of dropout biases the training to prioritize the earlier channels over the later ones.
Consequently, the encoder learns to encode more critical information (low frequency) in the initial feature maps and less important (high frequency) details in the subsequent ones; see Fig~\ref{fig:training}.
This prioritization enables flexible bitrate adaptation: upon encoding each frame, the encoder starts transmitting the most significant channels first.
Depending on the available bandwidth, the bitrate control module determines how many channels need to be sent to the decoder to ensure uninterrupted streaming; see Fig~\ref{fig:example}. Importantly, because the latent features are pre-ordered by significance, the bitrate control module basically acts like a simple gate and does not add any extra computational complexity to the encoder.

\section{Evaluation}
\label{sec:eval}
\vspace{-.5em}

\begin{figure}
\centering
\begin{subfigure}{.45\textwidth}
  \centering
  \includegraphics[width=.8\linewidth]{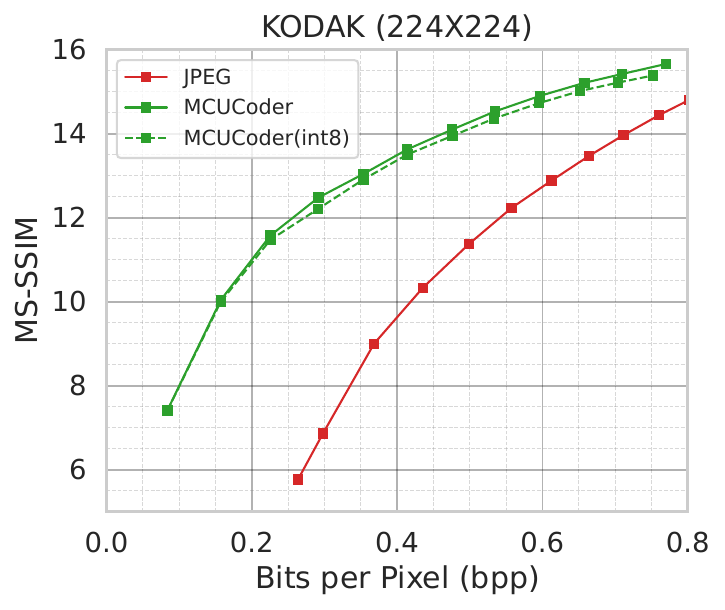}
  \label{fig:sub1}
\end{subfigure}%
\begin{subfigure}{.45\textwidth}
  \centering
  \includegraphics[width=.8\linewidth]{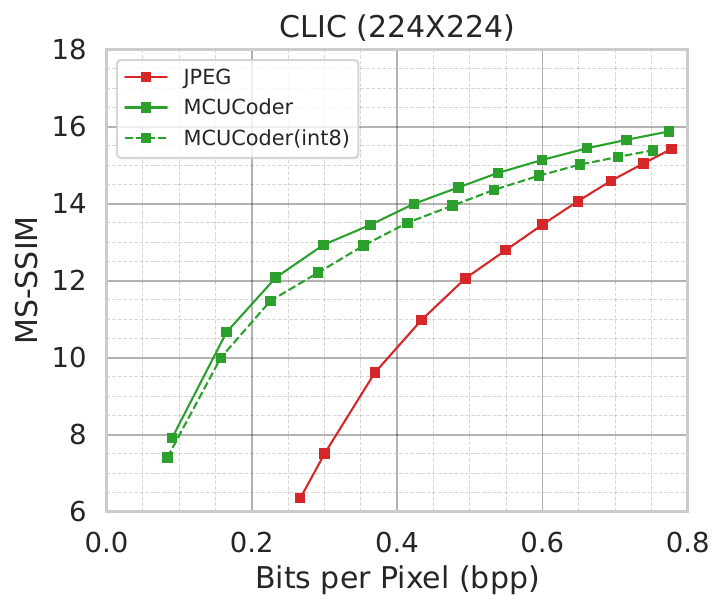}
  \label{fig:sub2}
\end{subfigure}
\begin{subfigure}{.45\textwidth}
  \centering
  \includegraphics[width=.8\linewidth]{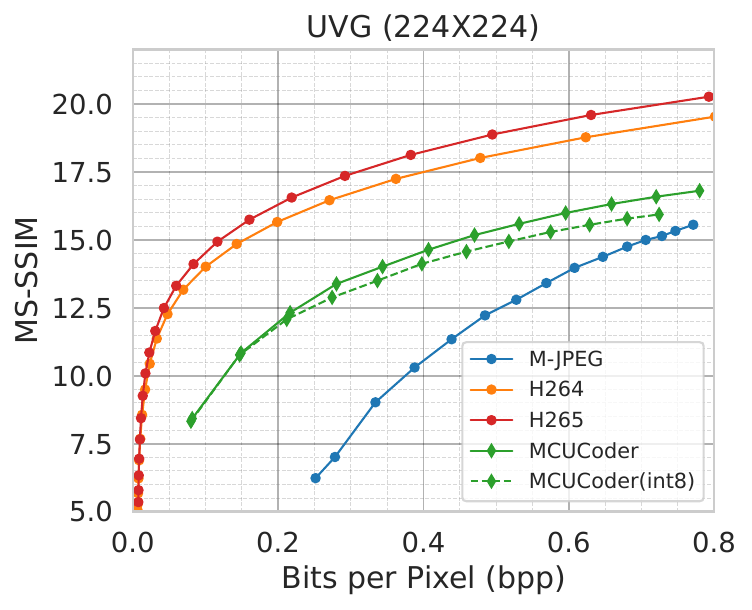}
  \label{fig:sub2}
\end{subfigure}
\begin{subfigure}{.45\textwidth}
  \centering
  \includegraphics[width=.8\linewidth]{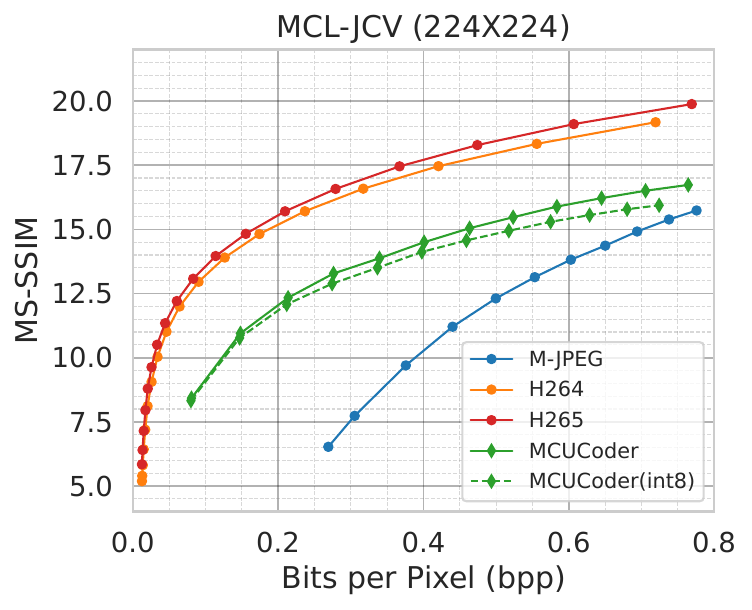}
  \label{fig:sub2}
\end{subfigure}
\caption{Comparison of \name (quantized and non-quantized model) and baselines on the image (KODAK \cite{kodak}, CLIC \cite{clic2020}) and video (MCL-JCV \cite{mcl_jcv}, UVG \cite{mercat2020uvg}) compression datasets. For context, we also compare with H.264 and H.265 on video datasets, despite being impractical for \glspl{MCU} due to high hardware demands. All datasets are resized to $224 \times 224$.}
\label{fig:msssim}
\vspace{-1em}
\end{figure}

We train \name on the 300K largest ImageNet images \cite{deng2009imagenet} and apply noise-downsampling preprocessing \cite{he2021checkerboard, balle2018variational}. We use Adam with an initial learning rate of $10^{-4}$ and a batch size of 16, and train for 1M iterations, lowering the learning rate to $10^{-5}$ in the final 50K iterations \cite{he2022elic}. To address quantization effects, we add random noise to the latent. Since \name is specifically designed for IoT environments, where the structure of the output is more critical than fine details, we use MS-SSIM as the loss function. We also quantize inputs, weights, and activations to INT8 for RAM efficiency and to leverage DSP and CMSIS-NN accelerators \cite{CMSIS-NN} in \glspl{MCU}. We use post-training quantization existing in TFLite-Micro \cite{tflite-micro} to reduce latency, processing power, and model size with minimal degradation in model accuracy. For all comparisons, we report performance metrics for both the FLOAT32 and INT8 models.

% cy and compatibility with DSP and CMSIS-NN accelerators.
% This process involved converting our PyTorch model to TensorFlow via ONNX, followed by further conversion to \texttt{model.cc} and \texttt{model.h}  files for deployment on MCUs.

\begin{figure*}[h] % Use figure environment to keep both aligned as figures
\centering
\begin{minipage}{0.35\textwidth} % Adjust the width as needed
    \centering
    \captionof{table}{\name (Quantized) BD-rate results. The anchor is \gls{M-JPEG}.}
    \label{tab:bdrate}
    \tiny
    \begin{tabular}{cccc}
        \toprule
        \multirow{2}{*}{\textbf{Type}} & \multirow{2}{*}{\textbf{Dataset}} & \multirow{2}{*}{\textbf{MS-SSIM}}  & \multirow{2}{*}{\textbf{PSNR}}  \\
         & & \\
        \midrule
        \multirow{2}{*}{Video} & MCL-JCV & -55.65\% & -47.39\% \\
         & UVG & -55.59\% & -35.28\% \\
        \midrule
        \multirow{2}{*}{Image} & KODAK & -55.75\% & -43.01\% \\
         & CLIC & -49.54\% & -38.02\% \\
        \bottomrule
    \end{tabular}
    \captionof{table}{Resource demands of \name on nRF5340 and STM32F7 \glspl{MCU}.}
    \label{tab:hardware}
    \tiny
    \begin{tabular}{lcc}
        \toprule
        & \textbf{nRF5340} & \textbf{STM32F7} \\
        \midrule
        Exec (ms) & 1,969 & 237 \\
        RAM (KB)  & 344 (33\%) & 360 (17\%) \\
        Flash (KB) & 100 (10\%) & 107 (5\%) \\
        \bottomrule
    % \vspace{-3em}
    \end{tabular}
    
    \label{bdrate_reversed}
\end{minipage}%
\hfill % Adds horizontal space between minipages
\begin{minipage}{0.55\textwidth} % Adjust the width as needed
    \centering
    \includegraphics[trim=60 10 20 20, width=0.6\textwidth]{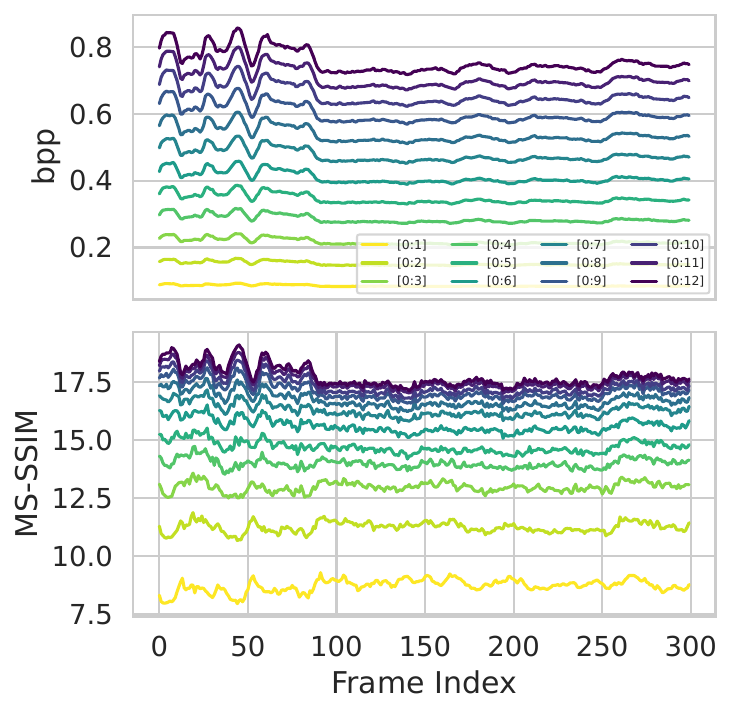}
    % \captionsetup{font=tiny}
    \caption{MS-SSIM and bpp for the SunBath video from UVG \cite{mercat2020uvg} dataset. [0:k] shows the use of the first k channels (out of 12) for decoding.}
    \label{fig:video_frame_sample}
    % \vspace{-2em}

\end{minipage}
    \vspace{-1em}

\end{figure*}

\subsection{Quantitative results}
\vspace{-.5em}
Due to the limited hardware resources of \glspl{MCU}, inter-frame compression is not practically feasible. As a result, in such devices, video compression is limited to \gls{M-JPEG} where each frame is compressed independently. Therefore, in addition to evaluating \name and its baselines from the perspective of video compression, we also assess its performance on image compression datasets. Given the lower resolution commonly encountered in IoT scenarios, we resize all the videos and images to $224 \times 224$.

% \begin{figure} % Use figure environment to keep both aligned as figures
% \centering
% \begin{minipage}{0.45\textwidth} % Adjust the width as needed
%     \centering
%     \captionof{table}{Resource requirements and inference time of \name on two \glspl{MCU}. The values in parentheses indicate the percentage of available hardware resources used .}
%     \label{tab:nrf}
%     \small
%     \begin{tabular}{lcc}
%         \toprule
%         & \textbf{nRF5340} & \textbf{STM32F7} \\
%         \midrule
%         Exec (ms) & 1,969 & 237 \\
%         RAM (KB)  & 344 (33\%) & 360 (17\%) \\
%         Flash (KB) & 100 (10\%) & 107 (5\%) \\
%         \bottomrule
%     \end{tabular}
% \end{minipage}%
% \hfill % Adds horizontal space between minipages
% \begin{minipage}{0.45\textwidth} % Adjust the width as needed
%     \centering
%     \includegraphics[width=\linewidth]{Pics/power.pdf}
%     \caption{Energy and current consumption of \name's encoder and JPEG for encoding one image on nRF5340.}
%     \label{fig:sub1}
%     \vspace{-3em}

% \end{minipage}
% \end{figure}

\textbf{Video compression:} We evaluate \name on the UVG \cite{mercat2020uvg} and MCL-JCV \cite{mcl_jcv} datasets, comparing its performance to \gls{M-JPEG} , see Fig.~\ref{fig:msssim}. For additional context, we include comparisons with traditional video codecs such as H.264 \cite{h264} and H.265 \cite{h265}, even though these codecs are impractical for deployment on \glspl{MCU} due to their significant computational and hardware demands.
Also, we report the Bjøntegaard Delta (BD) rate \cite{bjotegaard2001calculation} for both datasets in Table \ref{tab:bdrate}.
The results indicate that \name achieves a significantly higher MS-SSIM per bit compared to \gls{M-JPEG}, highlighting its ability to deliver better video quality at lower bitrates. This is especially valuable for IoT applications, where achieving high compression rates with minimal computational overhead is crucial due to limited hardware resources.
Additionally, \name has 12 "stacked" channels in its latent space, which provides 12 levels of quality that can be dynamically adjusted based on the available network bandwidth. In Fig.~\ref{fig:video_frame_sample}, we illustrate the bpp and MS-SSIM for each frame in a video from the UVG dataset for all 12 levels of quality. The results show that using more channels for decoding leads to a higher MS-SSIM, which verifies the effectiveness of the proposed stochastic dropout training. The PSNR results are reported in the Appendix \ref{appenix_C}.

\textbf{Image compression:}
To assess the image compression capabilities of \name, we conduct experiments on the CLIC \cite{clic2020} and KODAK \cite{kodak1993} datasets, see Fig.~\ref{fig:msssim}. The results in Table \ref{tab:bdrate} show that \name achieves an impressive average bitrate reduction of 55.75\% on the KODAK dataset and 49.54\% on the CLIC dataset, compared to JPEG. The PSNR results are reported in the Appendix \ref{appenix_B}.

\textbf{Latent ordering and DCT-JPEG alignment:} Fig.~\ref{fig:latent} shows the 12 latent channels obtained after training with the stochastic dropout method. The initial channels capture low-frequency information, while subsequent channels focus on high-frequency details. 
\begin{wrapfigure}{R}{5cm}
\includegraphics[trim=10 10 20 20, width=5cm]{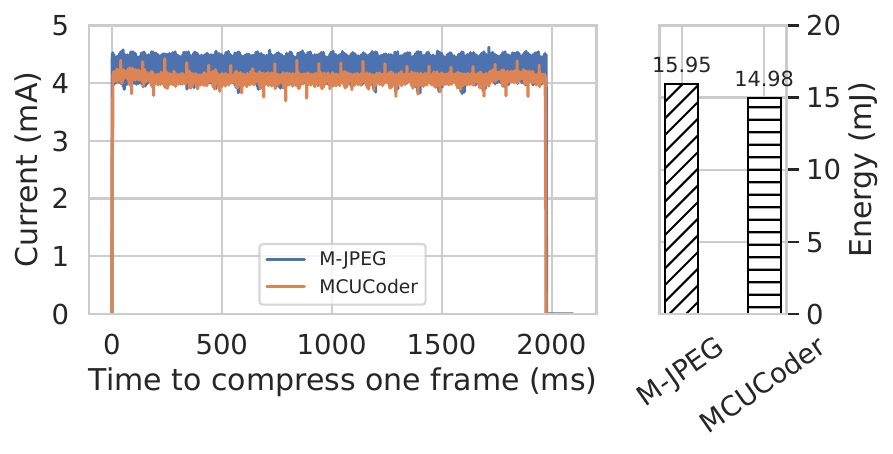}
\caption{Energy consumption of \name compared to \gls{M-JPEG}  for compressing one frame on the nRF5340.}
\vspace{-1em}
\label{fig:jpeg}
\end{wrapfigure}
Interestingly, this behavior mirrors the \gls{DCT} basis matrix employed in JPEG compression.

\textbf{Performance on MCUs}: We implement \name using TFLite-Micro \cite{tflite-micro} and Zephyr RTOS \cite{zephyr} on STM32F7 and nRF5340 \glspl{MCU}. The STM32F7 has 2 MB Flash, 2 MB RAM, and a Cortex-M7 processor, while the nRF5340 has 1 MB Flash, 512 KB RAM, and a Cortex-M33 processor, with both supporting DSP and CMSIS-NN acceleration \cite{CMSIS-NN}.
As reported in Table~\ref{tab:hardware}, \name uses 360 KB of RAM on the STM32F7 and 344 KB on the nRF5340, which is remarkably low and suitable for such constrained IoT devices.
To compare \name's energy consumption against \gls{M-JPEG}, we measured the energy consumption of \name and the optimized version of JPEG encoder for the Cortex-M series \cite{github:JPEGEncoder4Cortex-M} on the nRF5340, see Fig.~\ref{fig:jpeg}. The results indicate that \name matches JPEG’s energy consumption while significantly outperforming it in terms of BD-rate, see Table~\ref{fig:msssim}. The nRF5340 shows considerably slower performance than the STM32F7 for both \name and \gls{M-JPEG}, suggesting that it is better suited for event-driven applications rather than real-time streaming.

% (don't like this part)

% \begin{table}[h]
%     \small
%     \caption{Resource requirements and inference time of quantized \name int8 models without DSP acceleration on nRF5340 (Cortex-M33) and STM32F7 (Cortex-M7) MCUs.}
%     \centering
%     \begin{tabular}{lcccc}
%     \toprule

%     & \textbf{nRF w/o DSP} & \textbf{nRF w/ DSP} & \textbf{STM w/o DSP} & \textbf{STM w/ DSP} \\
%     \midrule
%     Exec (ms) & 45,590 & \textbf{1,969} & 4,882 & \textbf{237} \\
%     RAM (KB) & 344 ($33\%$) & \textbf{344 ($\textbf{33\%}$}) & \textbf{350 ($\textbf{17\%}$)} & 360 ($17\%$) \\
%     Flash (KB) & \textbf{94 ($\textbf{9\%}$}) & 100 ($10\%$) & \textbf{91 ($\textbf{4\%}$)} & 107 ($5\%$) \\
%     \bottomrule
%     \end{tabular}
%     \label{tab:nrf}
% \end{table}

% \begin{figure*}
%     \centering
%     \begin{subfigure}[b]{.8\textwidth}
%         \centering
%         \includegraphics[trim=0 0 0 0, clip, width=1\textwidth]{Pics/video_frame_sample.pdf}
%     \end{subfigure}
%     \caption{}  
% \label{fig:ARCH}
% \end{figure*}

% \begin{wrapfigure}{R}{3cm}
% \includegraphics[trim=20 10 20 20, width=3cm]{Pics/example.pdf}
% \caption{Stochastic dropout training}
% \vspace{-1em}
% \label{fig:HydraViT}
% \end{wrapfigure}
\section{Conclusion}
\vspace{-.5em}
\label{sec:conclusion}
We introduced \name, an ultra-lightweight asymmetric video compression model for resource-constrained IoT devices. With just 10.5K parameters and a 350KB memory footprint, compared to \gls{M-JPEG}, \name reduces bitrate by over 55\% on both the MCL-JCV and UVG datasets while matching the efficiency of \gls{M-JPEG}. Its adaptive bitrate streaming ensures smooth video transmission under fluctuating network conditions, making it ideal for edge applications.

% Paragraph option
% \paragraph{Paragraphs}

% inline citation
%  \citet{hasselmo}

\begin{ack}
This project has received funding from the Federal Ministry for Digital and Transport under the CAPTN-F\"{o}rde 5G project grant no.~45FGU139\_H and Federal Ministry for Economic Affairs and Climate Action under the Marispace-X project grant no.~68GX21002E.
\end{ack}

\bibliographystyle{unsrt}

\bibliography{BibTex}

%%%%%%%%%%%%%%%%%%%%%%%%%%%%%%%%%%%%%%%%%%%%%%%%%%%%%%%%%%%%

\appendix
\newpage
% \section{MS-SSIM on the KODAk dataset (original size)}
% \label{appenix_A}

% % \begin{figure}[!htbp]
% % \centering
% %   \centering
% %   \includegraphics[width=0.4\linewidth]{Pics/original_kodak_msssim.pdf}
% %   \caption{Comparison of \name (quantized and non-quantized model) and deep-baselines on the KODAK \cite{kodak} dataset. Due to the complexity and high RAM usage, none of the baselines are implementable on \glspl{MCU}.}
% %   \label{fig:sub1}
% % \end{figure}

\section{PSNR on KODAK and CLIC datasets}
\label{appenix_B}

\begin{figure}[!htbp]
\centering
\begin{subfigure}{.5\textwidth}
  \centering
  \includegraphics[width=0.8\linewidth]{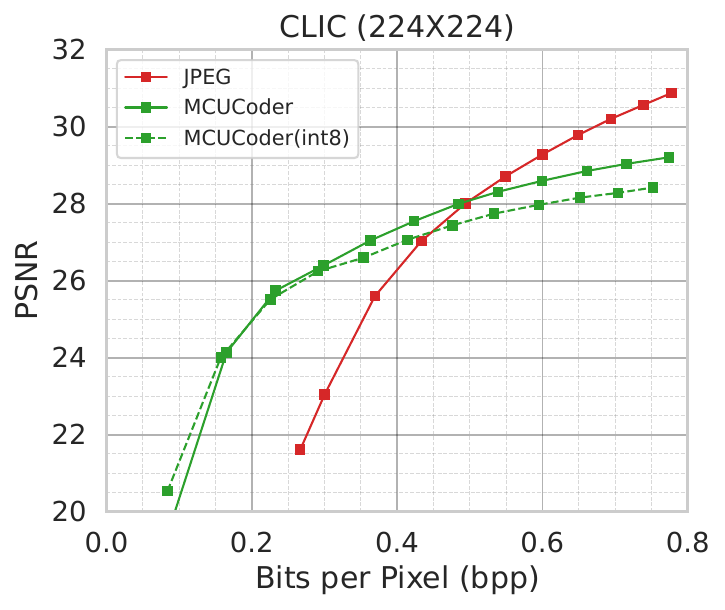}
  % \caption{A subfigure}
  \label{fig:sub1}
\end{subfigure}%
\begin{subfigure}{.5\textwidth}
  \centering
  \includegraphics[width=.8\linewidth]{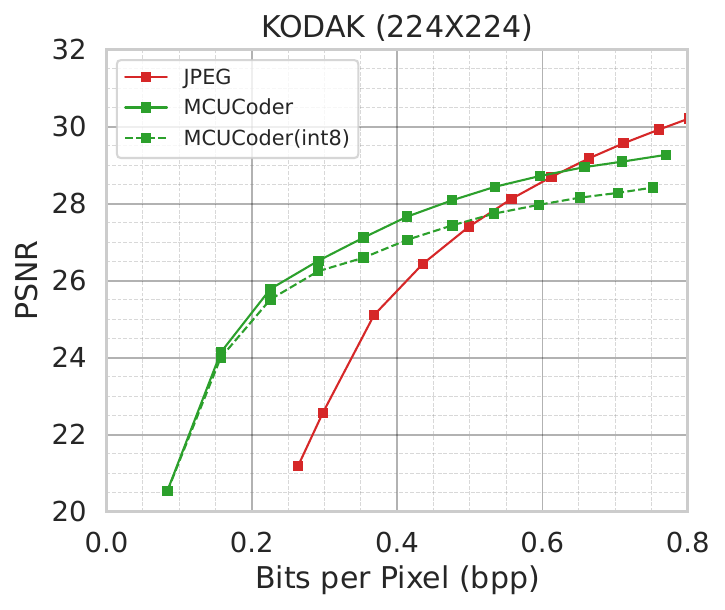}
  % \caption{A subfigure}
  \label{fig:sub2}
\end{subfigure}
\caption{Comparison of \name (quantized and non-quantized model) and baselines on the KODAK \cite{kodak} and CLIC \cite{clic2020} datasets. All datasets are resized to $224 \times 224$. Since \name is specifically designed for IoT environments—where structural integrity is prioritized over fine details—it has been optimized for MS-SSIM. Consequently, M-JPEG achieves better PSNR performance at higher bpp.}
\label{fig:test}
\vspace{-2em}
\end{figure}

\section{PSNR on the MCL-JCV and UVG datasets}
\label{appenix_C}

\begin{figure}[!htbp]
\centering
\begin{subfigure}{.5\textwidth}
  \centering
  \includegraphics[width=.8\linewidth]{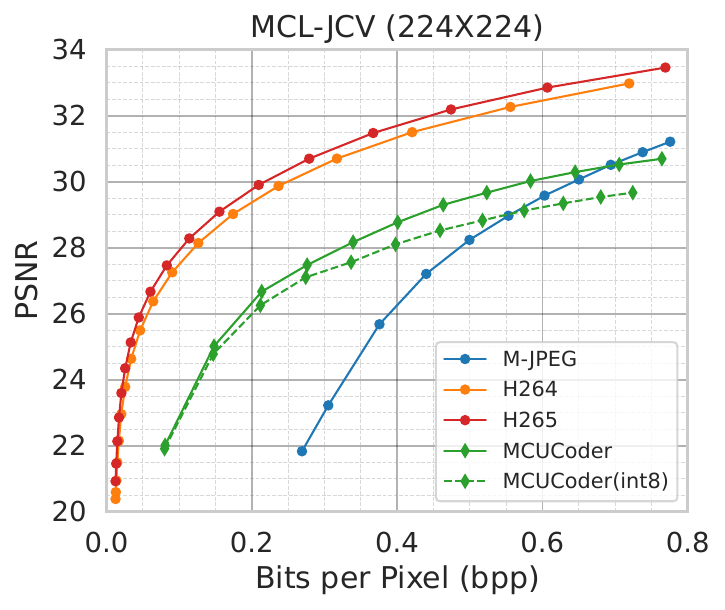}
  % \caption{A subfigure}
  \label{fig:sub1}
\end{subfigure}%
\begin{subfigure}{.5\textwidth}
  \centering
  \includegraphics[width=.8\linewidth]{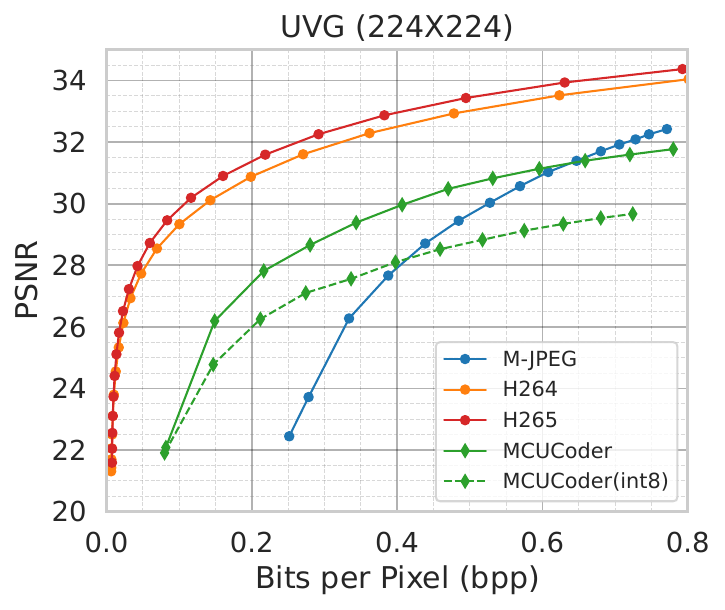}
  % \caption{A subfigure}
  \label{fig:sub2}
\end{subfigure}
\caption{Comparison of \name and baselines on MCL-JCV \cite{mcl_jcv} and UVG \cite{mercat2020uvg}. H.264 and H.265 are included for reference, though impractical for \glspl{MCU} due to hardware demands. \name, optimized for IoT with a focus on MS-SSIM, prioritizes structural integrity, while M-JPEG shows better PSNR at higher bpp.}

\label{fig:test}
\vspace{-2em}
\end{figure}

\section{Training logs of \name}
\label{appenix_D}
\vspace{-4em}

\begin{figure*}
    \centering
    \begin{subfigure}[b]{1.\textwidth}
        \centering
        \includegraphics[trim=0 0 0 0, clip, width=.85\textwidth]{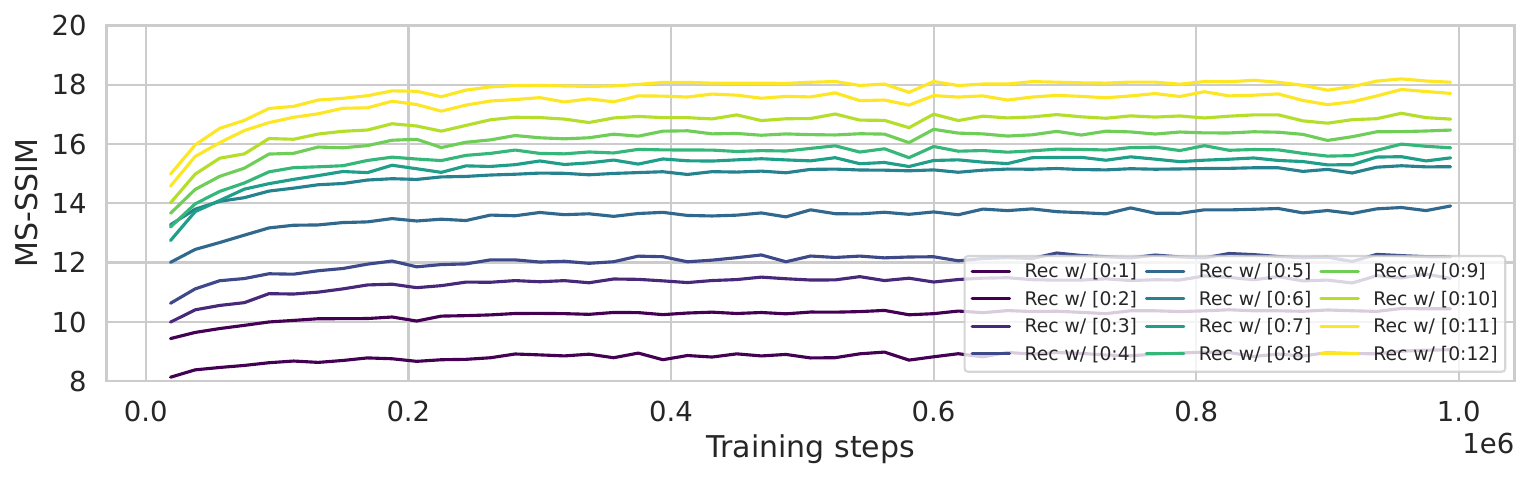}
    \end{subfigure}
    \caption{MS-SSIM values on the KODAK dataset during training. The notation $[0:k]$ represents the MS-SSIM of the reconstructed image using the first $k$ latent channels out of a total of 12. As shown, with stochastic dropout training, all the sub-latents can be trained simultaneously without overfitting to any particular sub-latent.}    
\label{fig:ARCH}
\end{figure*}

\newpage
\section{Examples of \name}
\label{appenix_D}

\begin{figure}[!htbp]
\centering
\begin{subfigure}{.45\textwidth}
  \centering
  \includegraphics[width=0.83\linewidth]{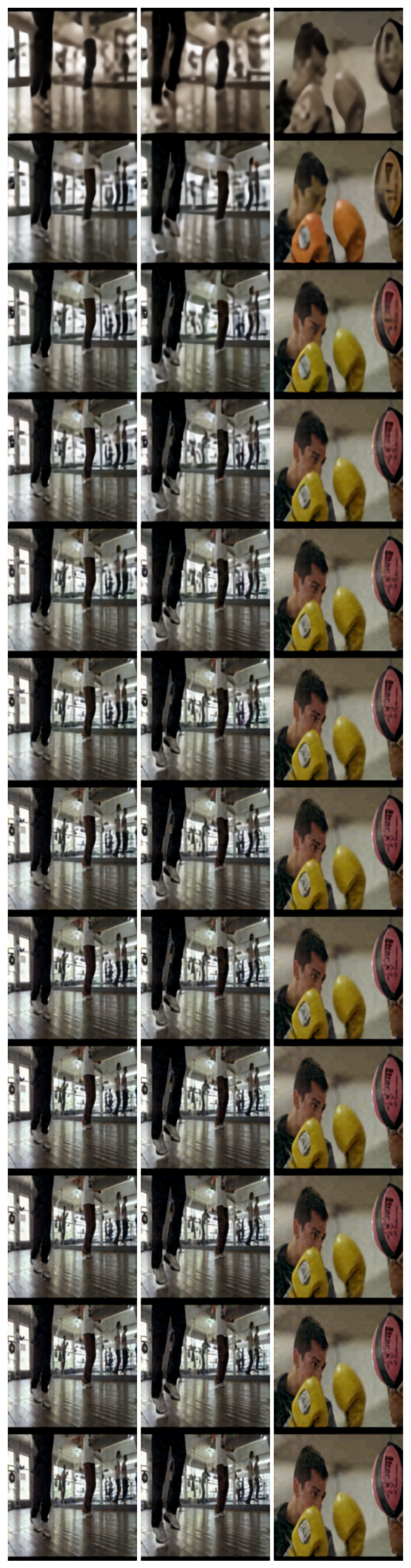}
  % \caption{A subfigure}
  \label{fig:sub1}
\end{subfigure}%
\begin{subfigure}{.45\textwidth}
  \centering
  \includegraphics[width=.83\linewidth]{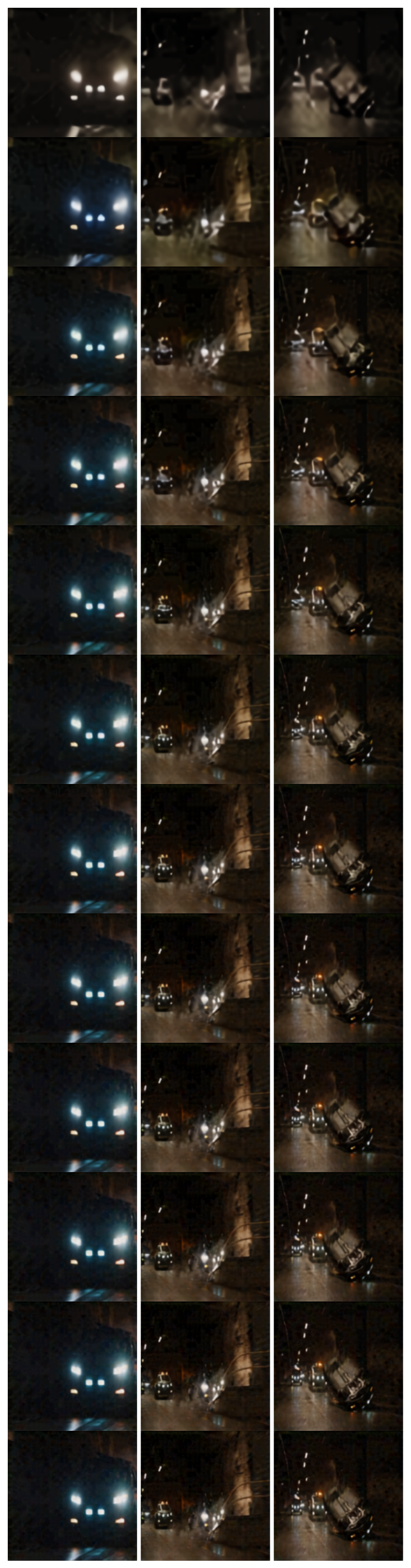}
  % \caption{A subfigure}
  \label{fig:sub2}
\end{subfigure}
\caption{Some samples from the MCL-JCV \cite{mcl_jcv} dataset. The columns represent different frames, while the rows display progressively improving levels of quality from top to bottom, produced by \name.}
\label{fig:test}
\end{figure}

\begin{figure}[!htbp]
\centering
\begin{subfigure}{.45\textwidth}
  \centering
  \includegraphics[width=.83\linewidth]{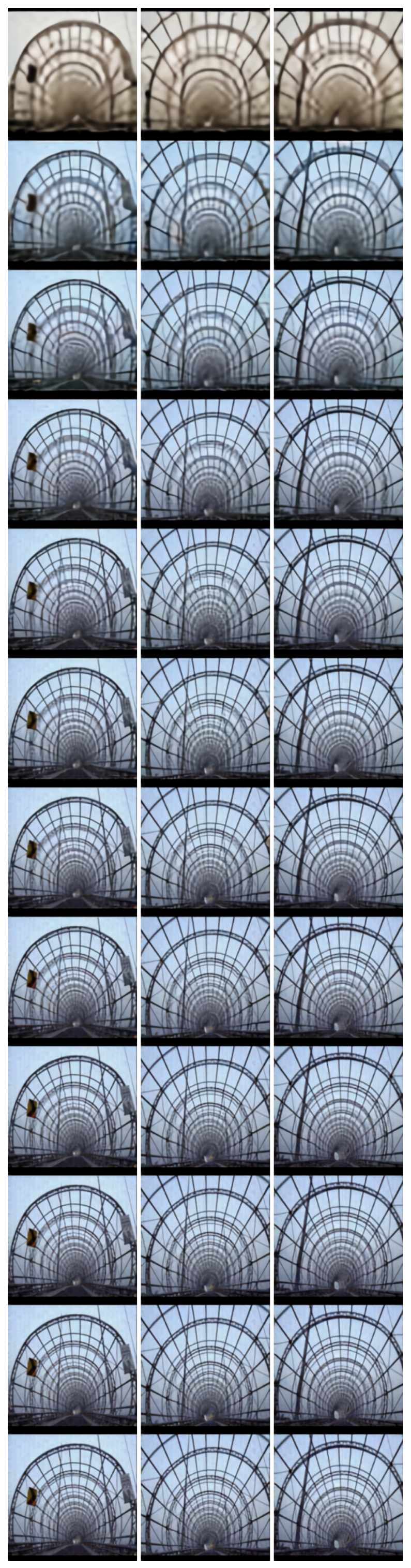}
  % \caption{A subfigure}
  \label{fig:sub1}
\end{subfigure}%
\begin{subfigure}{.45\textwidth}
  \centering
  \includegraphics[width=.83\linewidth]{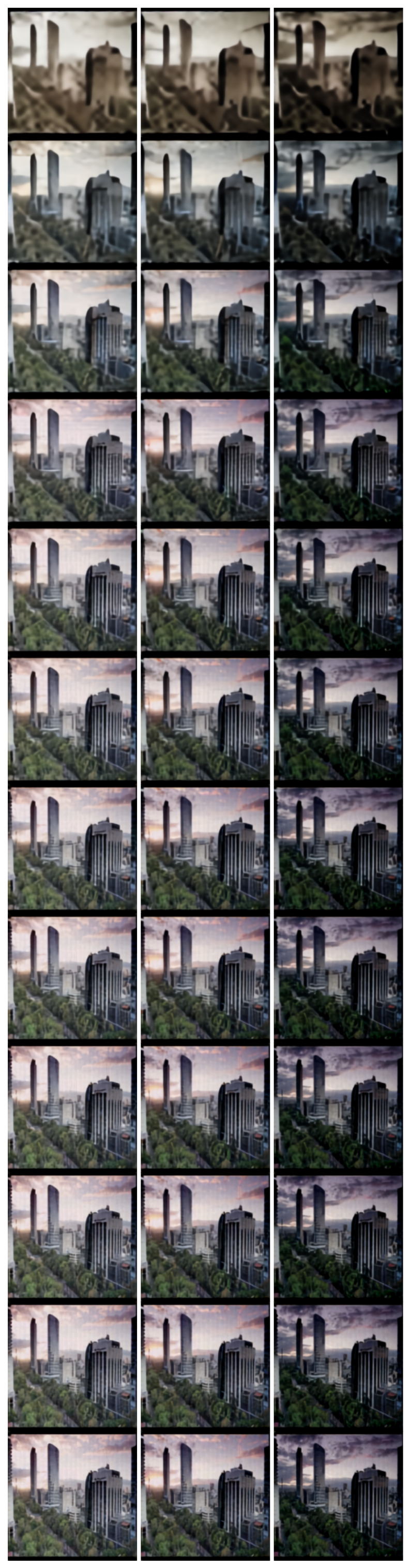}
  % \caption{A subfigure}
  \label{fig:sub2}
\end{subfigure}
\caption{Some samples from the MCL-JCV \cite{mcl_jcv} dataset. The columns represent different frames, while the rows display progressively improving levels of quality from top to bottom, produced by \name.}
\label{fig:test}
\end{figure}

%%%%%%%%%%%%%%%%%%%%%%%%%%%%%%%%%%%%%%%%%%%%%%%%%%%%%%%%%%%%

\end{document}